# A Novel Macroscopic Wave Geometric Effect of the Sunbeam and A Novel Simple Way to show the Earth-Self Rotation and Orbiting around the Sun

Sang Boo Nam
7735 Peters Pike, Dayton, OH 45414-1713 USA
sangboonam@mailaps.org

I present a novel macroscopic wave geometric effect of the sunbeam occurring when the sunbeam directional (shadow by a bar) angle c velocity is observed on the earth surface and a sunbeam global positioning device with a needle at the center of radial angle graph paper. The angle c velocity at sunrise or sunset is found to be same as the rotating rate of swing plane of Foucault pendulum, showing the earth-self rotation. The angle c velocity at noon is found to have an additional term resulted from a novel macroscopic wave geometric effect of the sunbeam. Observing the sunbeam direction same as the earth orbit radial direction, the inclination angle q of the earth rotation axis in relation to the sunbeam front plane is found to be related with the earth orbit angle, describing the earth orbit radial distance. The eccentricity of the earth orbit and a calendar counting days from perihelion are obtained by $dq/dt$ and q measured on the earth surface, showing the earth orbiting around the sun.

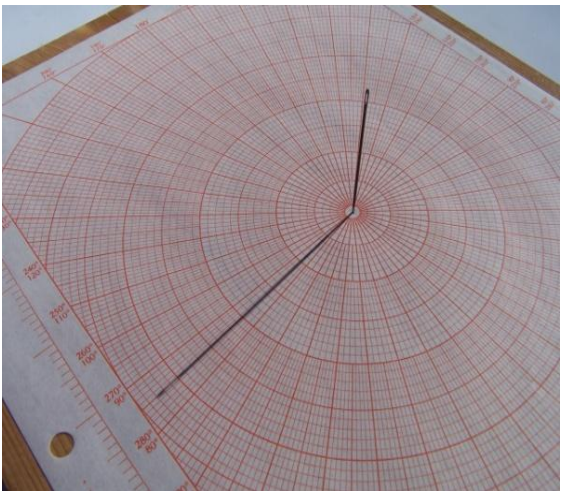

Fig. 1 Sunbeam Global Positioning Device with a needle at the center of radial angle graph paper.

Foucault demonstrated in 1851 the earth-self rotation (ESR) by showing the rotating rate of swing plane of Foucault pendulum [1] same as the projection of ESR angular velocity on the pendulum axis, and obtained the latitude at his location. But there is no way to show the earth orbiting around the sun (EOS) by Foucault pendulum. The sundial practices [2] from ancient to present are based on the sun appearing again in 24 hours later, but do not show ESR. The sundial data can be explained by the sun revolving around the earth once in 24 hours and the directional changes of the sun in seasons. Is there a way by which anybody can show ESR and EOS? Yes, I found one. The keys are the sunbeam directional angle c velocity and the velocity of the inclination angle q of the earth rotation axis in relation to the sunbeam front plane. The former is for ESR and the latter for EOS, respectively. It is natural to consider the angle c velocity for ESR, since the angle c on the earth surface is changing within a daytime. The angle c velocity at sunrise or sunset (AcVS) is found to be same as the rotating rate of swing plane of Foucault pendulum, showing ESR. But the angle c velocity at noon (AcVN) is found to be faster than AcVS. Both AcVS and AcVN have a contribution by the projection of the ESR angular velocity on the bar (a needle in Fig. 1) measuring sunbeam direction by its shadow, resulted from the macroscopic wave geometric effect (MWGE) [3] of the sunbeam. AcVN has an additional contribution by the projection of the ESR angular velocity on the shadow direction perpendicular to the bar. This counter intuitive effect is first time discussed here as a novel MWGE of the sunbeam, causing

AcVN faster than AcVS.   For EOS, realizing the sunbeam direction same as the earth orbit radial direction, I found a simple relation between the earth orbit angle R and the angle q accounting for lengths of day and night and seasons. The angle R velocity is obtained by the angle q velocity, and vice versa. Thus, the earth orbit radial distance is described by the angle q. Consequently, the eccentricity of the earth orbit and a calendar counting days from perihelion are obtained by dq/dt and q measured on the earth surface, contrast to Kepler's observations of celestial objects, planets.

   To study the sunbeam direction is equivalent to examining the shadow of a bar standing vertically on the earth surface as shown in Fig. 1. For the present study, we choose the earth center as the origin of coordinates as shown in Fig. 2. Let the earth-self rotating axis be the z-axis and the earth equatorial plane be the x-y plane. Let the z-y plane be the meridian, noon longitude, and b longitude measured from the meridian and u the latitude.  Let the sunbeam with propagation unit vector

$$\mathbf{k} = (0, -\cos q, \sin q)$$

come to the earth from far away, making angle q with the y-axis.

   We consider here the spherical earth. The sunset or sunrise longitude g can be determined by the crossing longitude with the sunbeam front plane (sfp), when sfp hits the earth center making angle q with the z-axis. Let us consider the latitude u circle of radius cosu at z = sinu in the unit of the earth radius. The distance from the z-axis to sfp on the z-y plane is sinu tanq equal to cosu cosg, the projection of the radius cosu on the z-y plane as shown in Fig. 2.  Thus, we obtain

$$\cos g = \tan q \tan u. \qquad (1)$$

This condition is for the radial unit vector

$$\mathbf{r}_g = (\cos u \sin g, \cos u \cos g, \sin u)$$

to be perpendicular to **k**, $\mathbf{k} \cdot \mathbf{r}_g = 0$.

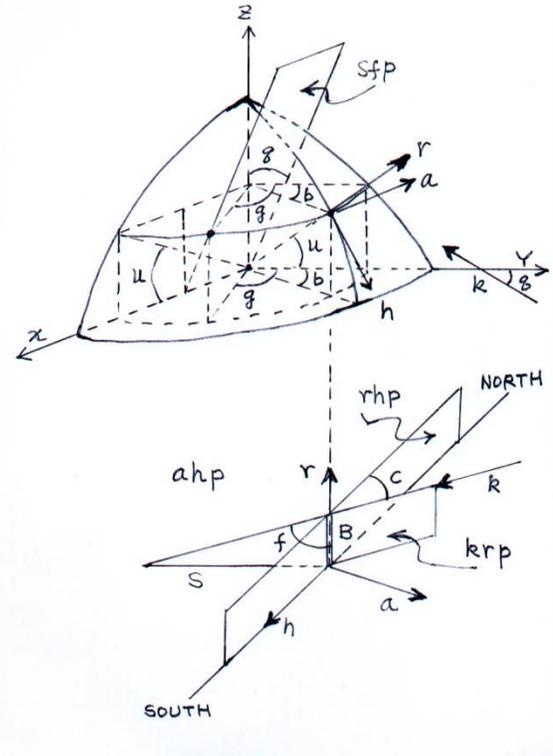

Fig. 2 sfp: sunbeam front plane, krp: **k-r** plane with the sunbeam and a bar, rhp: **r-h** longitudinal plane, ahp: **a-h** plane, angle f between the sunbeam and the bar, angle c between krp and rhp, u: latitude, longitudes b at the bar and g at sunset or sunrise measured from the z-y plane, the meridian, S: length of shadow on ahp, B: length of the bar stands vertically on ahp.

   For a given u, the lengths of day and night are determined by the angle q by Eq. (1). The g = 0 and g = π indicate no day and no night, respectively. By Eq. (1), the north pole region $|u| \geq \pi/2 - |q|$ has no day (no night) for positive (negative) q, and the south pole region the other way around.  At equinoxes, q = 0, g = π/2, we get equal day and night lengths. Strictly speaking, there is a small difference since q is changing from morning to sunset. At winter solstice, q is the earth axial tilt Q. The latitudes $|u| = \pi/2 - |Q|$ are called Arctic and Antarctic circles, respectively. The latitudes u = Q and u = − Q are known as Tropics of Cancer and Capricorn, respectively.

   At the place where a bar stands, the radial unit vector **r**, the azimuthal unit vector **a** and the longitudinal unit vector **h** are given by

**r = (cosu sinb, cosu cosb, sinu)**,
**a = (− cosb, sinb, 0)**, and
**h = (sinu sinb, sinu cosb, − cosu)**.

We then get the shadow length S by a bar with height B on **a-h** plane (ahp), with the angle f between the sunbeam and the bar, as

$$S = B \tan f, \quad (2)$$

$$\cos f = -\mathbf{k} \cdot \mathbf{r} = \cos u \cos q \cos b - \sin u \sin q. \quad (3)$$

The shadow angle c between **k-r** plane (krp) and **r-h** longitudinal plane (rhp), is given by

$$\cos c = \mathbf{a} \cdot (\mathbf{k} \times \mathbf{r}) / \sin f \quad (4)$$
$$= [\sin u \cos q \cos b + \cos u \sin q]/\sin f, \text{ or}$$
$$\tan c = \sin b/[\sin u \cos b + \cos u \tan q]. \quad (5)$$

For b = {0; g} = {Noon; Sunrise or Sunset}, we obtain the angles f and c as

$$f = \{u + q; \pi/2\}, \quad (6)$$
$$\tan f = \{\tan(u + q); \text{Infinite}\}, \quad (7)$$
$$c = \{0; \arccos[\sin q \sec u]\}. \quad (8)$$

By Eqs. (2), (6) and (7), we can examine the earth spherical nature by measuring the shadow lengths at two locations on the same longitudinal line. We expect two different shadow lengths would show the spherical earth. We can estimate the earth radius with the arc-distance between two locations and the difference between latitudes. At sunrise or sunset, from Eq. (8), the angle c is 90 degrees only at equinoxes, q = 0. In other words, sunrise and sunset directions are not always exactly east and west.

Let us examine angle f and c velocities. Taking derivatives of angles f by Eq. (3) and c by Eq. (5) with b, we obtain easily angle f and c velocities for b = {0; g} = {Noon; Sunrise of Sunset}, as

$$df/dt = BT \partial f/\partial b = \{0; \cos u \cos q \sin g\} BT, \quad (9)$$
$$dc/dt = BT \partial c/\partial b = \{\cos q \csc(u+q); \sin u\} BT. \quad (10)$$

Here t is the sidereal time and BT = $\partial b/\partial t$ = 15 deg/hr, the earth-self rotation angular velocity. There is no term of $\partial q/\partial b$ in Eq. (10) at noon. In other cases, it is negligibly small and neglected in Eq. (9) and Eq. (10). We estimate its size as following. The q changes 4x23.5 degrees in a year, $\partial q/\partial t$ = 94/365.25 deg/day = 0.01072 deg/hr and then $\partial q/\partial b$ = $(\partial q/\partial t)/(\partial b/\partial t)$ < 0.000715. Its omission is justified in the time frame of a daytime.

The angle c velocity at sunset or sunrise (AcVS) by Eq. (10) is given as

**15 sin(local latitude) deg/sidereal hour**. (11)

This is same as the rotating rate of swing plane of Foucault pendulum [1], showing the earth-self rotation. It is applied to measure the latitude 40.0 ± 1.0 degrees at my house, Dayton, OH, USA in a good agreement with 39.8 degrees given on the map. In practice, I taped a sheet of radial angle graph paper on a flat surface and put a needle vertically at the center of the graph paper as shown in Fig. 1. By reading the shadow angle c for every 5 or 10 minutes after sunrise about 30 minutes or so, AcVS was determined to get the latitude at our patio. The process was repeated near sunset.

The angle c velocity at noon (AcVN) by Eq. (10) is greater than AcVS. We see this by the ratio

$$AcVS/AcVN = \sin u \sin(u+q) \sec q$$
$$= 1 - (1 - \cos g) \cos^2 u \quad (12)$$

less than one. Why? To find the reason, we examine AcVN in BT unit. The $\partial c/\partial b$ in Eq. (10) at noon can be rewritten as

$$\cos q \csc(u+q) = \sin u + \cos u \cot(u + q). \quad (13)$$

The first term is resulted from the projection of BT on the bar. There is a precession vector [4]

$$\mathbf{G} = \cos u \csc(u + q) \quad (14)$$

in the sunbeam direction. Its projection on the shadow yields cosu, accounting for the projection of BT on the shadow. The projection of **G** on the bar yields the second term, cosu cot(u + q) in Eq. (13). There is no force involved. The first term in Eq. (13) is resulted from a MWGE of the sunbeam as the geometric effect of the rotating rate of swing plane of Foucault pendulum. The second term is a counter intuitive result, since it is resulted from the projection of BT on the shadow direction perpendicular to the bar. To my knowledge, this effect is first time shown as a novel MWGE of the sunbeam here. It becomes cotq at the equator, but vanishes at both poles, contrast to the first term whose values are zero at equator, 1 and – 1 at north and south poles, respectively. This can be realized as follows. The shadow is the sunbeam projected on ahp in Fig. 2 and its precession around the bar is the angle c velocity. Both projections of BT on the bar and shadow contribute to the angle c velocity. At sunset or sunrise, the projection of any vector in the sunbeam direction on the bar would vanish, since the sunbeam makes right angle with the bar. Thus, AcVS is obtained as given earlier on in Eq. (11), less than AcVN.

At the local noon, the shadow has the shortest length by Eq. (2) and AcVN the maximum value. The time difference between the local noon and standard time noon yields the local longitude = local standard longitude + BT x time difference hours. We now obtain the local latitude and longitude by the sunbeam global positioning device shown in Fig. 1. Later we show it can be used to obtain the earth orbit eccentricity and a calendar counting days from perihelion.

Moreover, we have found from Eq. (10) AcVN < BT for $\pi/2 - 2|q| < |u| < \pi/2 - |q|$, and AcVN > BT for $|u| < \pi/2 - 2|q|$.

We call the latitudes $|u| = \pi/2 - 2|Q|$ as Arctropic and Antarctropic circles where **AcVN = BT = 15 deg/hr**, respectively. They may be used as reference circles for longitudinal hour lines. In the region between reference circles, when $\partial c/\partial b = 1$, the angle c velocity = BT [5], it may be called the mid-morning and mid-afternoon, respectively.

To show EOS, as shown in Fig. 3, we derive a relation between the angle q and the earth orbit angle R measured from perihelion on the x-axis in the earth orbital system around S-sun.

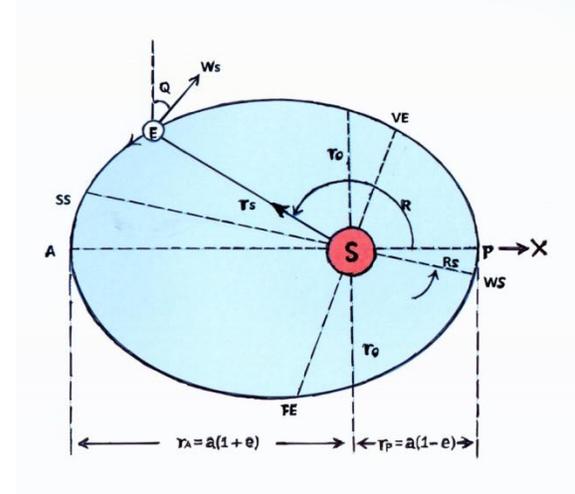

Fig. 3 Earth orbit radial distance r(R) = $r_0$/(1 + e cosR), R: the earth orbit angle, $r_0 = a(1 – e^2)$, a and e are the semi-major axis and eccentricity of the ellipse, P: perihelion, VE: vernal (spring) equinox SS: summer solstice, A: aphelion, FE: fall equinox, WS: winter solstice, $R_S$: angle at WS, Q: the earth axial tilt, $W_S$ = (sinQ cos$R_S$, sinQ sin$R_S$, cosQ): BT unit vector, $r_S$ = (cosR, sinR, 0): the r(R) unit vector.

The sunbeam unit vector in the earth system,

  **k = (0, - cosq, sinq) is $r_S$ = (cosR, sinR, 0),**

the earth orbit radial unit vector.
BT unit vectors in the earth and orbital systems are given by,

  **w = (0, 0, 1) and
  $w_S$ = (sinQ cos$R_S$, sinQ sin$R_S$, cosQ),**

respectively, where $R_S$ is the earth orbit angle R at winter solstice. Thus, we get easily [6]

  sinq = **k** • **w** = $r_S$ • $w_S$ = sinQ cos(R – $R_S$) or  (15q)

  R = $R_S$ + arc cos[sinq/sinQ].          (15R)

The $R_S$ is obtained by Eq. (19) with data of dq/dt at equinoxes.

By Eq. (15R), we get the earth orbit angle R with q and Q which are deduced from AcVN and the shadow length S at noon or AcVS and g at sunrise or sunset. Taking derivative of Eq. (15q) with time, we get the angle R velocity as

$$dR/dt = -(dq/dt) \cos q \csc Q \csc(R - R_S). \quad (16q)$$

This equals to that by the earth orbit [1]

$$dR/dt = L/M_E r^2(R)$$

$$= (L/M_E r_0^2)(1 + e \cos R)^2, \quad (16R)$$

$$r(R) = r_0/(1 + e \cos R), \quad (16r)$$

$$r_0 = a(1 - e^2) = L^2/M_E^2 GM_S,$$

$$L/M_E r_0^2 = 2\pi/Y(1 - e^2)^{3/2}, \text{ and}$$

$$Y = 2\pi a^{3/2}/(GM_S)^{1/2}. \quad (16Y)$$

Here Y is the Kepler period of the elliptic orbit [7], L the earth orbital angular momentum, $M_E$ the earth mass, $M_S$ the mass of sun, G the Newton gravitational constant, r(R) the earth orbit radial distance from the sun, e and a are the earth orbit eccentricity and the semi-major axis of the earth orbit, respectively.

Let us consider the function Z(R) defined, for $|R| \leq \pi/2$, by

$$Z(R) = [(dR/dt)(R)/(dR/dt)(R + \pi)]^{1/2}$$

$$= r(R + \pi)/r(R) \quad (17r)$$

$$= (1 + e \cos R)/(1 - e \cos R) \quad (17R)$$

$$= |(dq/dt)(R)/(dq/dt)(R + \pi)|^{1/2}. \quad (17q)$$

Z(R) has a maximum value at R = 0, which can be easily found. By Eq. (17R) and Eq. (17q), we get

$$e = [Z(0) - 1]/[Z(0) + 1]. \quad (18)$$

The eccentricity of the earth orbit is obtained with data of dq/dt at perihelion and aphelion.

Moreover, by Eq. (15q) and Eq. (16q), dq/dt has local extreme values at equinoxes (q = 0), which can be easily found. By Eq. (17R) and Eq. (17q) for $Z(R_S + \pi/2) = Z_E$, we obtain

$$R_S = \arcsin[(1 - Z_E)/e(Z_E + 1)]. \quad (19)$$

The angle $R_S$ at winter solstice is obtained with data of dq/dt at equinoxes. By integrating dR/dt of Eq. (16R), we obtain a calendar, D days from perihelion as

$$D = (Y/2\pi)[H - eE^{1/2} \sin R/(1 + e \cos R)], \quad (20)$$

$$H = 2 \arctan[K \tan(R/2)],$$

$$E = 1 - e^2, \text{ and}$$

$$K = [(1 - e)/(1 + e)]^{1/2}.$$

R and q are related by Eq. (15q) and Eq. (15R). The second term in Eq. (20) shows the earth elliptic orbit. For e = 0, D = (Y/2π) R is obtained for the circular orbit. We demonstrate numerically EOS by making a seasonal calendar from Eq. (20) as follows.

Let us consider e = 0.01671 [say by Eq. (18)] and $R_S$ = -10 degrees on 21Dec [by Eq. (19)] and the period of q = Y = 365.2425 days [7]. R angles at vernal equinox, summer solstice and fall equinox are given by $R_S + \pi/2$, $R_S + \pi$ and $R_S + 3\pi/2$, respectively. Inserting the above into Eq. (20), we get 89.06810 days from winter solstice to vernal equinox (89 days to 20Mar), 92.87826 days to summer solstice (93 days to 21June), 93.56980 days to fall equinox (94 days to 23Sept), 89.72634 days back to winter solstice (89 days to 21Dec), respectively. Calculated days of seasons are in good agreements with real seasonal durations given in above brackets.

For a given D days from perihelion, R is obtained by Eq. (20), then q by Eq. (15q) and the sunrise or sunset g by Eq. (1). The lengths of day and night are predicted. In a simple term, EOS causes the changes of lengths of day and night as well as seasonal durations.

To collect optimally the solar energy, the solar panel should be directly toward to the sun. It should in the morning be vertically to almost east, having the angle c at sunrise, and rotate, accordingly to equations for f and c angles, to the middle sky simultaneously inclined angle f to have u + q at noon. Afternoon, it reverses its process, to west.

It is worthy to stress again a novel MWGE of the sunbeam causes AcVN faster than AcVS. The sunbeam through a pinhole complimentary to the shadow would yield same results.

**Photon has so many beautiful sides** [8].

I sincerely thank Professor Sung Ho Choh for communicating to me his data of the angle c velocity around noon. Using Eq. (10) with u = 37 degrees, the calculated AcVN 24.93 and 15.80 deg/hr for q = 0 and q = 23.5 degrees are in good agreements with his data 25.0 ± 1.0 and 15.3 ± 0.5 deg/hr measured on 2008/3/21 and 2006/12/23, respectively.